\documentclass[pre,superscriptaddress,twocolumn,a4paper,showpacs]{revtex4-1}
\usepackage{amssymb}
\usepackage{wasysym}
\usepackage{graphicx}
\usepackage{epsfig}
\usepackage{subfigure}
\usepackage{float}
\usepackage{xcolor}
\begin{document}

\title{Mobility helps  problem-solving systems to avoid Groupthink}

\author{Paulo F. Gomes}
\affiliation{Instituto de F\'{\i}sica de S\~ao Carlos,
  Universidade de S\~ao Paulo,
  Caixa Postal 369, 13560-970 S\~ao Carlos, S\~ao Paulo, Brazil}
\affiliation{Instituto de Ci\^encias Exatas e Tecnol\'ogicas, Universidade Federal de Goi\'as, 
75801-615  Jata\'{\i}, Goi\'as, Brazil }

\author{Sandro M. Reia}
\affiliation{Instituto de F\'{\i}sica de S\~ao Carlos,
  Universidade de S\~ao Paulo,
  Caixa Postal 369, 13560-970 S\~ao Carlos, S\~ao Paulo, Brazil}
               
 \author{Francisco A. Rodrigues}
 \affiliation{Instituto de Ci\^encias Matem\'aticas e de Computa\c{c}\~ao,
              Universidade de S\~ao Paulo,
             Caixa Postal 668, 13560-970 S\~ao Carlos, S\~ao Paulo, Brazil}
\affiliation{Mathematics Institute, University of Warwick, Gibbet Hill Road, Coventry CV4 7AL, UK}
\affiliation{Centre for Complexity Science, University of Warwick, Coventry CV4 7AL, UK}
 
\author{Jos\'e F.  Fontanari}
\affiliation{Instituto de F\'{\i}sica de S\~ao Carlos,
  Universidade de S\~ao Paulo,
  Caixa Postal 369, 13560-970 S\~ao Carlos, S\~ao Paulo, Brazil}

\begin{abstract}
Groupthink occurs when everyone in a group starts thinking alike, as when people put unlimited faith in a  leader. Avoiding this phenomenon is a ubiquitous challenge to problem-solving enterprises and typical countermeasures  involve the  mobility  of  group members.
  Here we  use an agent-based model of imitative learning to study  the influence of the mobility  of the agents on the time they  require to find the global maxima of   NK-fitness landscapes.   
The agents cooperate by exchanging   information on their fitness and use
this information  to copy the fittest agent in their influence neighborhoods,  which are  determined by  face-to-face interaction networks. The influence neighborhoods are variable since  the  agents perform  random walks  in a two-dimensional space.
We find that mobility is slightly harmful for solving easy problems, i.e. problems that do not exhibit suboptimal solutions or local maxima. For difficult problems, however, mobility can prevent the imitative search being trapped in suboptimal solutions  and guarantees a better performance than the independent search for any system size.
\end{abstract}

\maketitle

\section{Introduction}\label{sec:intro}

Learning through observation and imitation are central   to the success of the human species as they are  key elements to the construction of culture  \cite{Boyd_05,Rendell_10,Perc_17}. In the context of collective intelligence, this significance is neatly  summarized  in the  phrase ``Imitative learning acts like a synapse, allowing information to leap the gap from one creature to another'' \cite{Bloom_01}. Imitation begets the question of who should be imitated, a decision that was probably shaped by natural selection   and impacted greatly on the social organization and behavioral patterns of gregarious animals \cite{Blackmore_00}. 

It has been hinted  that agent-based models of  imitative learning could reproduce some features of  the problem-solving  performance of task forces \cite{Fontanari_14,Fontanari_15}. In fact, for a variety of combinatorial optimization problems,  a search procedure based on  imitative learning  yields a substantial  improvement on the group performance as compared with the independent search, where  the agents explore the  solution space  of the problem independently of each other, provided that the imitation propensity of the agents and the group size are set  to appropriate values.    However,  if  the agents are too willing to imitate their more successful peers or the group is too large,  then the imitative learning search   yields a calamitous performance which  is reminiscent of   the Groupthink phenomenon of social psychology that occurs when everyone in a group starts  thinking alike \cite{Janis_82}.

Groupthink and the consequent entrapment in suboptimal solutions poses a  hard challenge to problem-solving enterprises in general. In the academic world, for instance, this issue is  tackled by either calling for outside experts or allowing sabbatical leaves to group members.  Here we examine if this remedy to Groupthink, namely,  the mobility of agents,   works  for the  agent-based model of imitative learning  too.

More pointedly,
we carry out extensive Monte Carlo simulations of  systems of mobile agents that  use imitative learning to  search for the global maxima of  NK-fitness landscapes \cite{Kauffman_87}. The agents exchange information on their fitness and  imitate the fittest  agent -- the model -- in their influence neighborhoods, which are  determined by  face-to-face interaction networks \cite{Zhao_11}.   Data on the physical proximity and face-to-face contacts of individuals in numerous real-world situations were recorded by the SocioPatterns collaboration \cite{socio} and used to study  general aspects of human behavior  \cite{Cattuto_10,Starnini_12,Starnini_13}  as well as the patterns of transmission of infectious diseases in human populations \cite{Sapienza_18,Moinet_18}. In face-to-face networks, the agents interact (i.e. imitate  the models)  if the distance between them is less than some prespecified threshold.
In addition, the agents  move in a square box  by performing steps of fixed length in random directions in the plane.

We find that mobility is slightly detrimental in the case of easy problems, i.e. additive landscapes with a single maximum, for which imitation of the model agents is guarantee of getting closer to the solution of the problem, i.e. the global maximum. In this case, strengthening the spatial and fitness  correlations of agents in closed gatherings  yields the optimal problem-solving performance. However, for difficult problems, i.e. rugged landscapes with many local maxima (suboptimal solutions), mobility can prevent the imitative search being trapped in the local maxima and guarantees a better performance than the independent search for any system size. This finding is all the more remarkable because mobility does not change the topological properties of the underlying  face-to-face network, such as the typical number of agents within an influence neighborhood, and so its beneficial effect on the system performance  is purely dynamical and cannot be achieved
through the rewiring of the links between agents \cite{Perra_12}.

The effects of mobility have been considered for a variety of  collective phenomena such  as  the synchronization of chaotic oscillators \cite{Frasca_08,Fujiwara_11}, the emergence of cooperation in evolutionary game theory \cite{Meloni_09,Cardillo_09} and disease spreading \cite{Colizza_06,Buscarino_08}, to mention only a few. These works assert that  moderate mobility  can promote the emergence of synchronization and cooperation, whereas high mobility can disrupt those collective behaviors. Moreover, in the context of disease spreading, mobility can significantly reduce the epidemic threshold. Hence, here   we follow  the established practice of statistical physics and complexity science of studying the effects of  mobility on the emergent and collective properties of individual-based models by addressing its effects on
 the performance of cooperative problem-solving systems.

The rest of this paper is organized as follows. In Section \ref{sec:NK} we outline the NK model of rugged fitness landscapes  \cite{Kauffman_87} which we use to represent the optimization problems the  agents must solve.
 The  random motion in the two-dimensional physical space where the agents are placed and the imitative search on the state space of the NK model  are explained in Section \ref{sec:model}.  
In  Section \ref{sec:res} we present and analyze the results of our simulations, emphasizing the influence of the mobility of the agents on the 
problem-solving performance of the imitative search.  Finally, Section \ref{sec:disc} is reserved to our concluding remarks.


\section{NK-fitness landscapes}\label{sec:NK}

The  agents must find  the unique global maximum of  a fitness landscape generated using   the
NK model  \cite{Kauffman_87}.  Although this model was originally proposed to explore optimization principles in population genetics and developmental biology, its influence has gone far  beyond the biological realm  \cite{Kauffman_95} and  the NK model is now the paradigm  for problem representation in management research  \cite{Levinthal_97,Lazer_07,Billinger_13},
as it allows the  tuning  of the ruggedness of the fitness landscape  and hence of the difficulty of the problem.  

The NK-fitness landscape is defined in the space of binary strings of length $N$  and so this parameter determines the size of the solution or state space, namely, $2^N$.  The other parameter  $K =0, \ldots, N-1$  determines the range of the epistatic interactions among the bits of the binary string and  influences strongly the number of local maxima on the landscape. We recall that two bits are said to be epistatic whenever the combined effects of their contributions to the fitness of  the binary string  are not  merely additive \cite{Kauffman_87}. More pointedly, for each string $\mathbf{x} = \left ( x_1, x_2, \ldots,x_N \right )$ with $x_i = 0,1$ we associate a fitness value
\begin{equation}\label{NK_1}
\mathcal{F} \left ( \mathbf{x}  \right ) = \frac{1}{N} \sum_{i=1}^N f_i \left (  \mathbf{x}  \right ) ,
\end{equation}
where $ f_i$ is the contribution of component $i$ to the  fitness of string $ \mathbf{x}$. It is assumed that the functions  $ f_i$ are $N$ distinct real-valued functions on $\left \{ 0,1 \right \}^{K+1}$ that  depend on the state
$x_i$  as well as on the states of the $K$ right neighbors of $i$, i.e., $f_i = f_i \left ( x_i, x_{i+1}, \ldots, x_{i+K} \right )$ with the arithmetic in the subscripts done modulo $N$. As usual,  we assign to each $ f_i$ a uniformly distributed random number  in the unit interval \cite{Kauffman_87}. Because of the randomness of $f_i$, we can guarantee that  $\mathcal{F}  \in \left ( 0, 1 \right )$ has a unique global maximum and that different strings have different fitness values. We recall that  a string is a maximum if its fitness is greater than the fitness of 
all its $N$ neighboring  strings (i.e., strings that differ from it at a single bit).

For $K=0$ the landscape has a single maximum, which is easily determined by picking for each component $i$ the state $x_i = 0$ if  $f_i \left ( 0 \right ) >  f_i \left ( 1 \right )$ or the state  $x_i = 1$, otherwise. In addition, this landscape is clearly additive since the fitness of a string is completely determined by the sum of the components $f_i  \left ( x_i \right )$. The other extreme $K=N-1$ results in landscapes in which the fitness of neighboring strings are uncorrelated and so, in this case, the NK model  reduces to the Random Energy model \cite{Derrida_81}. This uncorrelated landscape   has on the average  $2^N/\left ( N + 1 \right)$ maxima with respect to single bit flips \cite{Kauffman_87}. We note that  for $K>0$ finding the global maximum of the NK model is an NP-complete problem \cite{Solow_00}.

Since our goal is to study the effects of the mobility of the agents on the performance of  the imitative search,  we must guarantee that the agents  explore the same  fitness landscape. Distinct landscape realizations generated using the same  values of the parameters $N$ and $K>0$ may differ greatly  in their numbers  of local maxima. 
Therefore, here we fix the string length  to $N=12$  and consider a single realization of  a rugged landscape with degree of epistasis $K=4$. This rugged landscape has $50$ maxima.  The fitness of the global maximum is $\mathcal{F}_{gm} \approx 0.783$,  the average fitness of the local maxima is $\bar{\mathcal{F}}_{lm} \approx 0.682$,  and the  average fitness of the landscape is  $\bar{\mathcal{F}} \approx 0.508$. 
We consider, in addition, a smooth landscape ($K=0$) that allows us to single out the influence of the local maxima on the performance of the imitative search.
The  small size of the state space ($2^{12} = 4096$ binary strings of length $12$)  enables the full exploration of the space of parameters and, in particular, the study of the regime where the time required to find the global maximum is much greater than the size of   the solution space.

\section{Model}\label{sec:model}

\begin{figure} 
\centering  
 \includegraphics[width=0.48\textwidth]{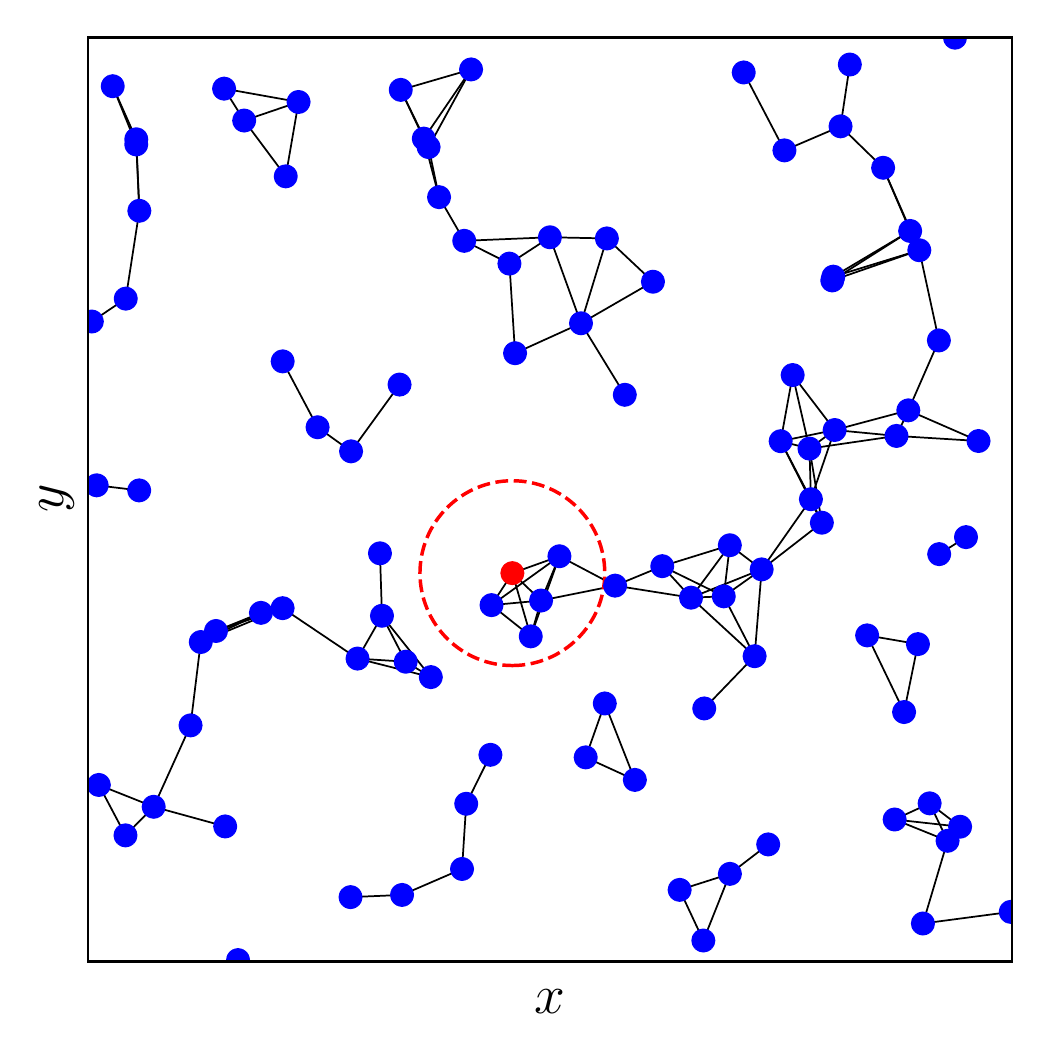}  
\caption{Snapshot of a system of $M=100$ agents and density $\rho = 0.0512$.  Agents within a distance $d = \alpha  d_0 $,  where $d_0 =  1/\sqrt{\rho}$ and  $\alpha=1$,  are connected by a link.  The dashed circle of radius $d$ centered at the target agent determines its influence neighborhood, which comprises four agents in this example. Links  that cross the square box borders are not shown for  sake of clarity.  
 }  
\label{fig:1}  
\end{figure}

We consider a system of $M$ agents placed  in a   square box of linear size $L$ with periodic boundary conditions (i.e., a torus).  In the initial configuration, the coordinates $x$ and $y$ of each agent are chosen  randomly and uniformly over the  length $L$. The density of agents
$\rho = M/L^2$,  which we fix to $\rho = 0.0512$ throughout the paper, yields the relevant spatial scale to analyze the  motion of the agents on the square box. In fact, since the effective  area of an agent is $1/\rho$, the  quantity $d_0 = 1/\sqrt{\rho}$ can be viewed as the  linear size or, for short, the size of an agent and it will be our standard  to measure all distances in our study.    More pointedly, we measure the distance $d$ within which interactions between agents are allowed in units of $d_0$, i.e., $d = \alpha d_0$ with $\alpha > 0$. The set of agents inside a circle of radius $d$ centered at a particular agent constitutes the influence neighborhood from where it will select a model to imitate. This scenario is illustrated in Fig.\ \ref{fig:1} that shows a snapshot of  a system of  $M=100$ agents in the square box.  Henceforth we refer to the network created  by the union of the  influence neighborhoods  as the influence network. This is the classic random geometric graph  originally introduced to model wireless communication networks \cite{Gilbert_61} and that  was recently  used  as a face-to-face network  in the  modeling of the  dynamics of human interactions \cite{Starnini_13}. We note that the fixed value of  the density $\rho$ is inconsequential, provided we use $d_0$ as the standard for measuring distances in the square box.

Each agent is represented by a binary string  of length $N$, whose bits  are  initially   drawn at  random with equal probability for $0$ and $1$.  The agents explore the NK-fitness landscape aiming at finding its global maximum  by flipping bits  following the rules of the imitative learning search \cite{Fontanari_15},  which consist basically of copying a bit of the  fittest agent in their influence neighborhoods as will be described in detail in this section.  In addition, the agents  move randomly around the square box, thus changing their influence neighborhoods and, in principle, affecting the  efficiency of the imitative search. Next we describe the movement in the  $2^N$-dimensional space of the binary strings and the physical motion on the square box. Henceforth we will  use  the terms agent and string interchangeably.

The dynamics begins with the selection of an agent at random, the so-called target agent,  at time $t=0$ and  comprises two stages. The first stage is the motion on the square box: an angle $\theta \in \left [ 0, 2 \pi \right )$  is chosen randomly to give the direction of motion   and then a fixed step of length $\delta d_0$ with $\delta \geq 0$ is taken on that direction. Once the target agent is at the new position, a circle of radius $d = \alpha d_0$  is drawn around it so that its influence neighborhood is determined, as shown in Fig.\ \ref{fig:1}. Then the second stage, namely, the update of the string of the target agent (or the target string, for simplicity) sets in. If the influence neighborhood is empty, i.e. there is no agent within a distance $d$ from the target agent, or all agents in the
influence neighborhood   have fitness lower than  or equal to the fitness of the target agent, then the target agent  simply flips a  bit  at random. Note that due to the nature of the NK-fitness landscape, two agents that have the same fitness must be identical (clones).

A more interesting situation is when there are agents with fitness higher than the fitness of the target agent in its influence neighborhood. Then there are two possibilities of action. The first, which happens with probability $1-p$,   consists of flipping a  bit  at random of the target string as before.    The second, which happens with probability $p$, is the  imitation  of a model string, which is the string  of highest fitness  in the  influence  neighborhood of the target agent. In this case, the  model and the target  strings are compared  and the different bits are singled out.  Then the target agent selects at random one of the distinct bits and flips it so that this bit is now the same in both strings.   Hence,  imitation  results in  the increase of the similarity between the target and the model agents, which may not necessarily lead to  an increase of the fitness of the target agent if the landscape is not additive, i.e. for $K > 0$.

The parameter $p \in \left [0,1 \right ]$ is the imitation probability, which we assume is  the same for all agents (see \cite{Fontanari_16} for the relaxation of this assumption). The case $p=0$ corresponds to the baseline situation in which    the agents   explore the state space independently of each other and so, in this case, the motion on the square box has no effect at all on the performance of the search. The case $p=1$ corresponds to the situation where only the model strings explore the state space through  random bit flips, whereas the other strings simply follow the models in their influence neighborhoods.  
The  imitation   procedure described above was borrowed from the incremental assimilation mechanism \cite{Shibanai_01,Avella_10,Peres_11} used to study the influence of external agencies  in  the celebrated Axelrod's model  of social influence   \cite{Axelrod_97}.

After the target agent is updated, which means   performing a step of size $\delta d_0$ in a random direction and flipping a bit  of its string,   we increment the time $t$ by the quantity $\Delta t = 1/M$.   Then another agent is selected at random and the procedure described above is repeated. Note that during the increment from $t$ to $t+1$, exactly  $M$ moves and string operations are performed, though not necessarily by $M$ distinct agents. This asynchronous update  seems more appropriate to  simulate the  continuous-time motion  of the agents, as well as  the  bit changes in the strings. Use of an alternative synchronous update would introduce a global clock that has no counterpart in the problem we seek to model.

The  search ends when one of the agents finds the global maximum and we denote by $t^*$ the halting time. The efficiency of the  search is measured by the total number of string operations necessary  to find that maximum, i.e. $Mt^*$,  
and so the  computational cost of the search is  defined as 
\begin{equation}\label{C}
C = M t^*/2^N ,
\end{equation}
 where, for convenience, we have rescaled $t^*$ by the size of the state space $2^N$.  To aid the understanding of the model, in the Appendix we offer a probabilistic description of the states of the agents and derive a master equation for the imitative search.

\section{Results}\label{sec:res}

To evaluate the  performance of the imitative search we use, as usual, the mean  computational cost $\langle C \rangle $, which  is obtained by averaging the computational cost $C$ over $10^5$  searches for the same  landscape realization. Since  our main concern here is  the  spatial  distribution and motion of the agents on the square box, we will fix  the imitation probability to $p=0.5$, and focus on  the system size   $M$ as well as on the parameters $\alpha$ and $\delta$ that specify the radius of the influence neighborhood and the step size of the random motion of the agents, respectively.

\begin{figure}
\centering  
 \subfigure{\includegraphics[width=0.48\textwidth]{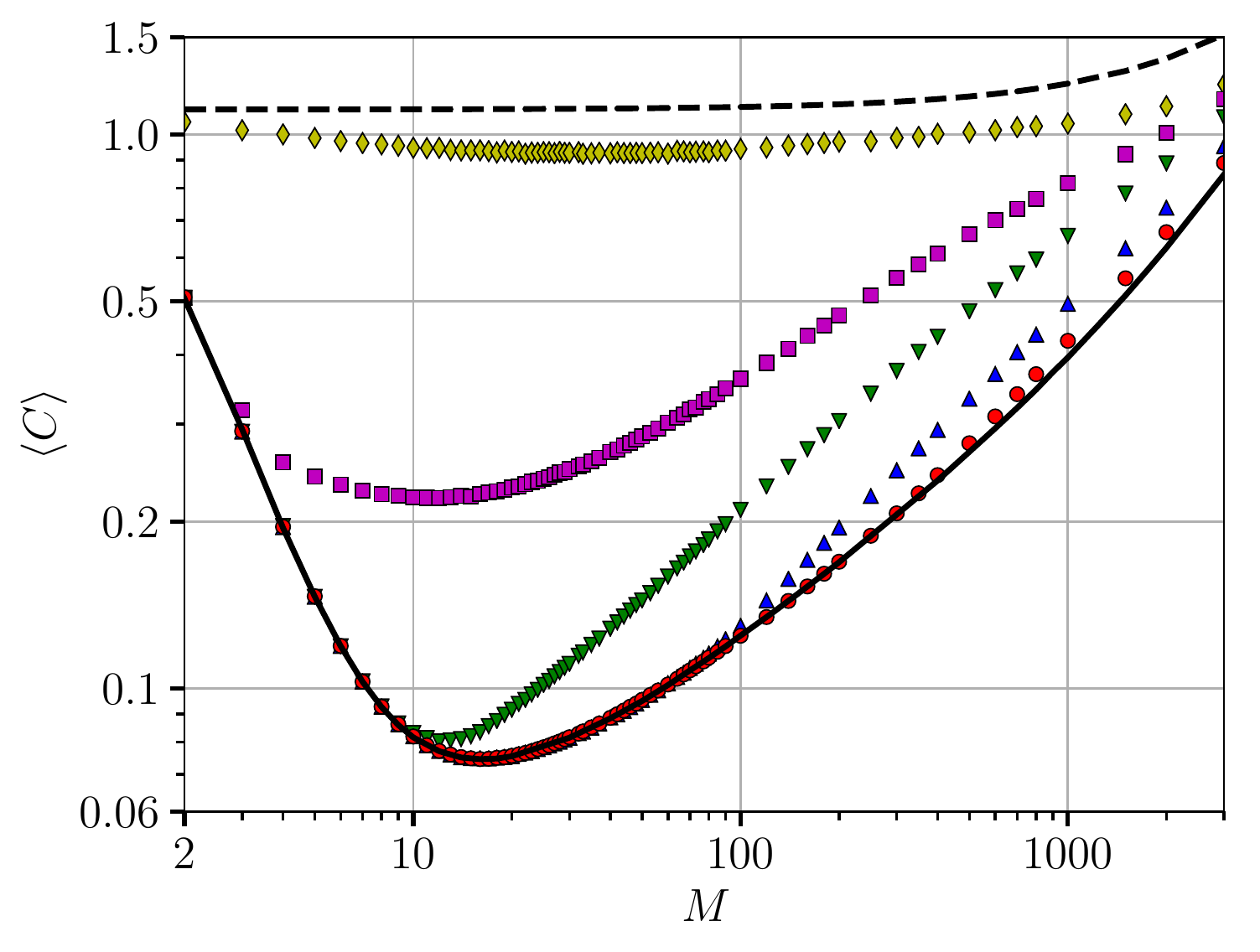}} 
 \subfigure{\includegraphics[width=0.48\textwidth]{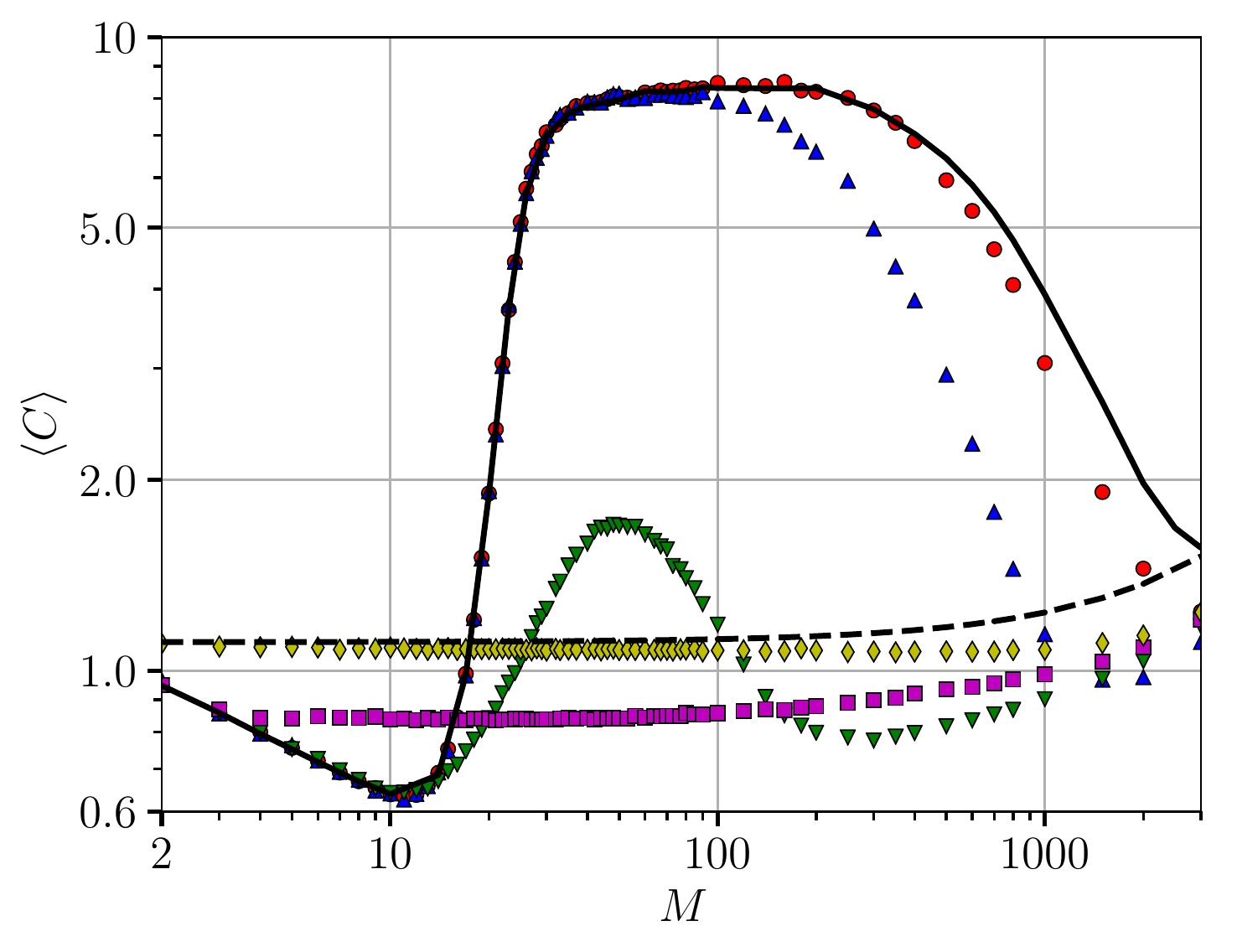}}
\caption{Mean computational cost $\langle C \rangle $ as function of  the system size $M$ for the position-fixed scenario ($\delta=0$). The  imitation probability is $p=0.5$ and the  radius of the influence neighborhood is $d = \alpha  d_0 $ with  $\alpha= 0.25 (\blacklozenge), 1 (\blacksquare), 2  (\blacktriangledown), 5 (\blacktriangle) $ and $10 (\CIRCLE)$.    The solid curve is the result for the fully connected system and the dashed curve is the analytical  result for the independent search \cite{Fontanari_15}.
The upper panel shows the results for the smooth landscape ($K=0$) and the lower panel for the rugged landscape ($K=4$).
 }  
\label{fig:2}  
\end{figure}

\subsection{Position-fixed  scenario}

To better appreciate the influence of the mobility on the performance of the imitative search, we study first the case where the agents remain fixed at their (random) positions specified in the initial set up of the system.  This position-fixed scenario (i.e. $\delta=0$) is  useful for understanding the role of the parameter $\alpha$ that determines the radius of the influence neighborhood, $d = \alpha d_0 $, where $d_0$ is the linear size of the agent.

\begin{figure}  
\centering  
 \includegraphics[width=0.48\textwidth]{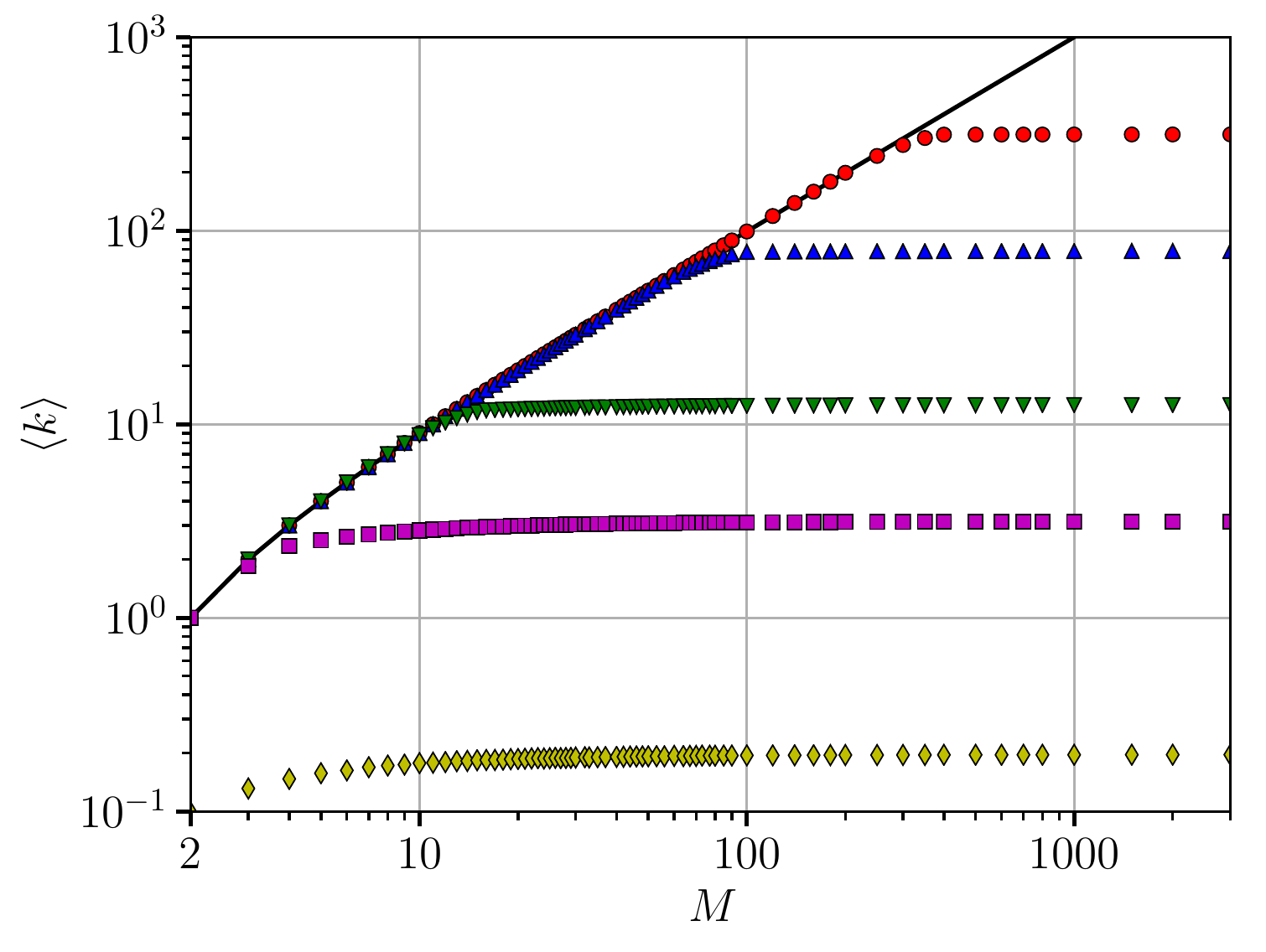}  
\caption{Average connectivity $\langle k \rangle $ of the influence network as function of  the system size $M$ for the position-fixed scenario  ($\delta=0$). The radius of the influence neighborhood is $d = \alpha  d_0 $ with  $\alpha= 0.25 (\blacklozenge), 1 (\blacksquare), 2  (\blacktriangledown), 5 (\blacktriangle) $ and $10 (\CIRCLE)$. The solid curve is the result for the fully connected network, $\langle k \rangle = M-1$. For  $M \to \infty$ we have $ \langle k \rangle \to \pi \alpha^2$.
}  
\label{fig:3}  
\end{figure}

Figure \ref{fig:2} shows that the effect of 
$\alpha$ on the  mean computational cost  depends on the ruggedness of the landscape.  As expected,  when  $\alpha < 1$, so that the interaction distance is less than the size of the agent $d_0$, the agents  explore the state space practically independent of each other and the computational cost is essentially the same as the cost of the independent search, which is very little sensitive to changes on  $M$, provided that $M \ll 2^N$ (see \cite{Fontanari_15} for the analytical derivation of the computational cost of the independent search). The parameter  $\alpha$ correlates strongly with  the average connectivity   $\langle k \rangle$ of
the influence network, which  is solely determined by the influence neighborhoods of the agents and by the system size $M$, as shown in Fig.\ \ref{fig:3}. This correlation  explains the effect of $\alpha$ on the performance of the search. In fact, since for the  smooth landscape ($K=0$) the fitness of the agents offer reliable information about their  distances to the global maximum, expanding the influence neighborhoods while keeping $M$  fixed  increases the odds of  finding a high fitness  model agent, which then  boosts the system performance. In this case, the best performance is achieved by a fully connected system, which is  shown in Fig.\ \ref{fig:2} as a solid curve for clarity purposes, although the results were also obtained through simulations.

The scenario changes drastically for the rugged landscape ($K=4$) due to the presence of local maxima whose main detrimental effect is to uncouple the fitness of an agent  from its distance to the global maximum.  As a result,  agents at local maxima spread unreliable information to their followers  that   may trap the entire system in a suboptimal solution.  The  catastrophic performance  observed in the case  of 
densely connected networks and large system sizes is akin to the Groupthink phenomenon  \cite{Janis_82},  when everyone in a group starts thinking alike,  which can occur when people put unlimited faith in a  leader (the model agent). A way of circumventing  Groupthink   is to limit or delay the flow of information among the agents and this can be achieved by reducing their influence neighborhood  or, equivalently, the average connectivity of the influence network \cite{Rodrigues_16} (see Fig.\ \ref{fig:3}). There is, however, a tradeoff  between avoiding the local maximum traps and optimizing the search performance.  For instance, the choice $\alpha =1$ (see  lower panel of Fig.\ \ref{fig:2})  avoids those traps altogether and always yields a superior performance compared with the independent search, but  it  misses the optimal performance that can be  achieved for larger values of $\alpha$ at $M \approx 12$.
These large values of $\alpha$, however, expand the  influence neighborhoods thus making the system much more susceptible to  Groupthink as  shown in Fig.\ \ref{fig:2}.  

Large systems  increase the attractivity of the local maxima, thus producing the undesired Groupthink,  because they allow the existence of several copies of the model agent in a same influence neighborhood. Although the model agent can escape the local maximum by flipping a bit at random according to the rules of the  imitative search, the extra copies will quickly attract the updated model agent back to the local maximum, resulting in the very high computational costs shown in Fig.\ \ref{fig:2} for the rugged landscape. For the smooth landscape, however, a large system size results in an increased computational cost  (though it is always smaller than the cost  of the independent search) simply because of duplication of work since only $M \approx N = 12$ agents are necessary to explore the neighborhood of the model string and thus to find a fitter string that is closer to the global maximum.

For small system sizes $M$ and $\alpha$ not too small, the results of Fig.\ \ref{fig:3} (and of Fig.\ \ref{fig:2} as well) show that the system is fully connected, i.e., $\langle k \rangle = M-1$. This happens because the density $\rho$ (and hence $d_0$) is kept constant  so that when  $M$ changes, the linear size $L$ of the square box changes too, while the radius of the interaction neighborhood $ d = \alpha d_0$ 
remains the same. As a result,  for small $M$ (and small $L$) the interaction neighborhood of an agent is likely to comprise the entire square box. In the other extreme, $M \to \infty$ (and hence $L \to \infty$) the agents are uniformly distributed over the square box and so  the average number of agents inside the interaction neighborhood  of area $\pi d^2$ is simply $ M \pi d^2/L^2 = \pi \alpha^2$, in agreement with the results of Fig.\ \ref{fig:3}.

 \subsection{Mobile-agents  scenario}
 
 We turn now to the more interesting situation where the agents move in random directions with a step of fixed length $\delta d_0$. The obvious effect of this motion is to make the influence neighborhood of the agents volatile, but the manner this volatility influences the performance of the  imitative search  is far from obvious as we will see next.
 
\begin{figure}
\centering  
 \subfigure{\includegraphics[width=0.48\textwidth]{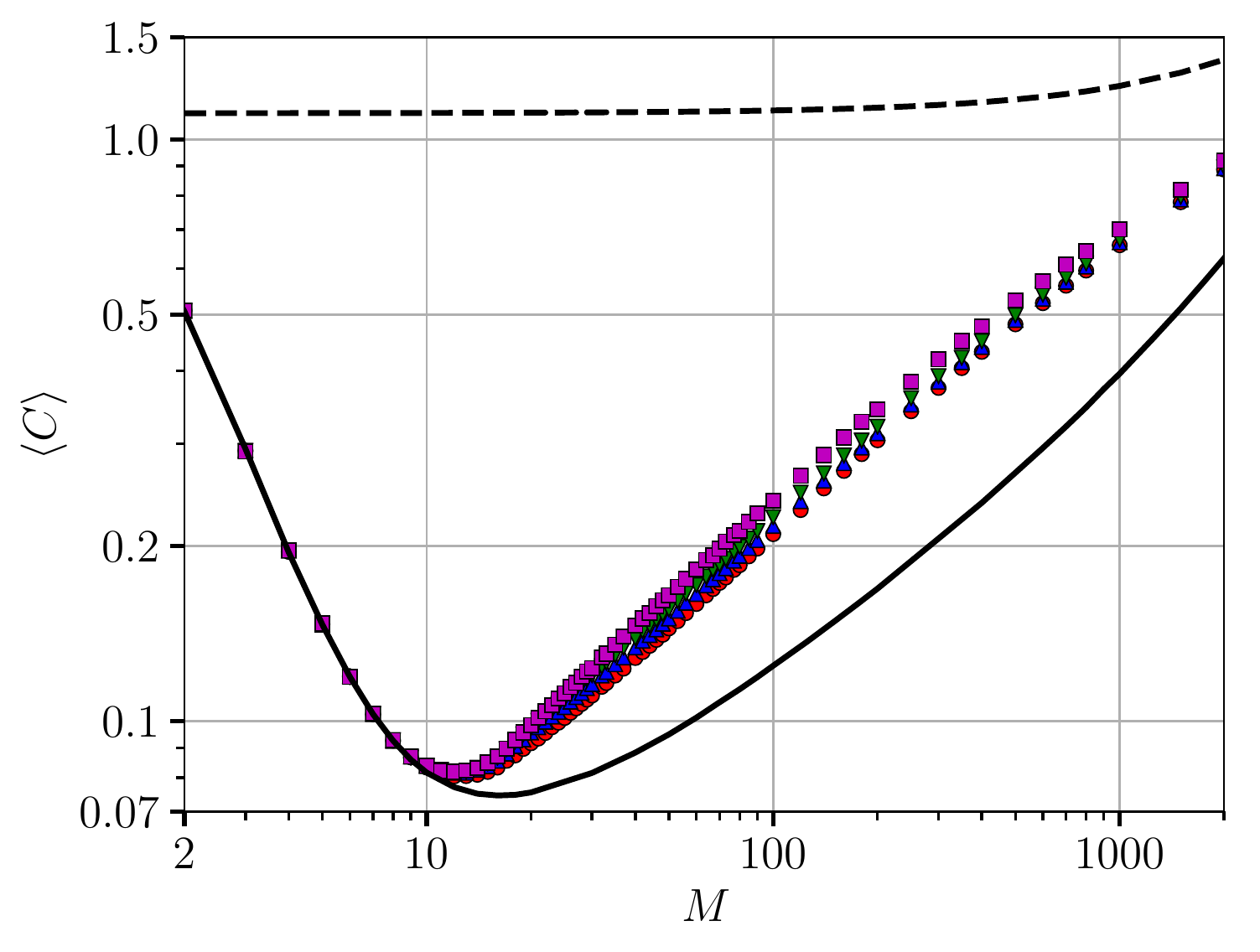}} 
 \subfigure{\includegraphics[width=0.48\textwidth]{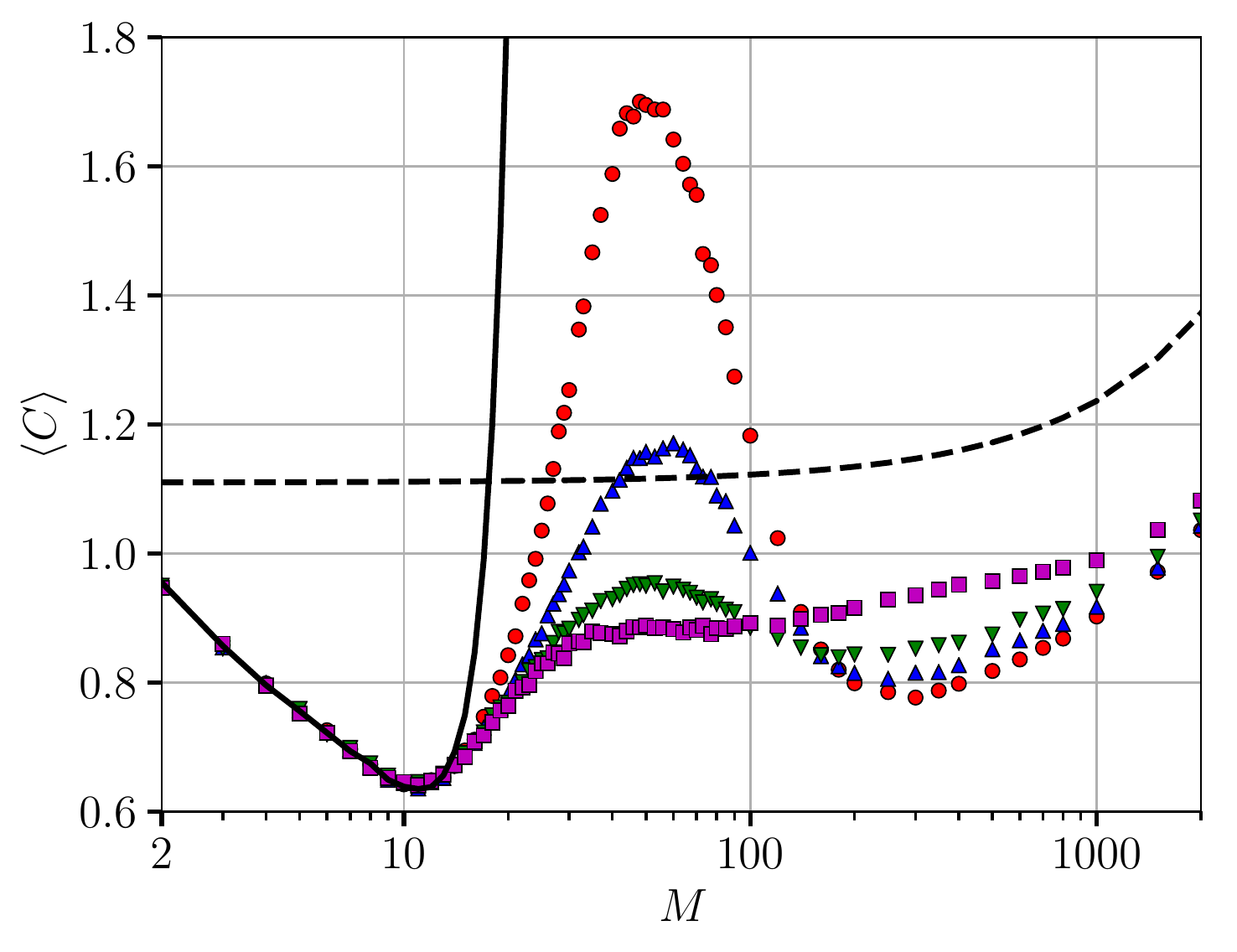}}
\caption{Mean computational cost $\langle C \rangle $ as function of  the system size $M$ for mobile agents with step sizes $\delta = 0 (\CIRCLE), 0.5 (\blacktriangle), 1 (\blacktriangledown)$ and $ 100 (\blacksquare)$. The  imitation probability is $p=0.5$ and the  radius of the influence neighborhood is $d = \alpha  d_0 $ with  $\alpha=  2$. The solid curve is the result for the fully connected system and the  dashed curve is the analytical  result for the independent search \cite{Fontanari_15}.
The upper panel shows the results for the smooth landscape ($K=0$) and the lower panel for the rugged landscape ($K=4$).
 }  
\label{fig:4}  
\end{figure}
 
 Figure  \ref{fig:4} shows the influence of the mobility on the computational cost for our two fitness landscapes.  A nonzero step size $\delta$  produces only a mild degradation on the performance of the search for the smooth landscape ($K=0$)  and so the effect of the mobility in this case is hardly noticeable. For the rugged landscape ($K=4$), however,  the mobility  is very effective in avoiding the traps of the local maxima  without the incurred tradeoff observed in the position-fixed  scenario. This is so because the mobility does not change the average connectivity of the influence network. In fact,  measurement of the mean  degree of  the agents by averaging over all the configurations in a  run and then averaging over distinct runs yields the same results obtained for  the position-fixed scenario (see Fig.\ \ref{fig:3}). Hence, the random motion of the agents does not alter the nature of the influence  network. This observation makes the results of Fig.\  \ref{fig:4} even more remarkable since it reveals that  the change in the computational cost is a genuine effect of the mobility of the agents and not a consequence of  changing the connectivity of   the  influence network.  Of course,  if the system is fully connected, then  the mobility will not affect the performance of the search.

\begin{figure}
\centering  
 \includegraphics[width=0.48\textwidth]{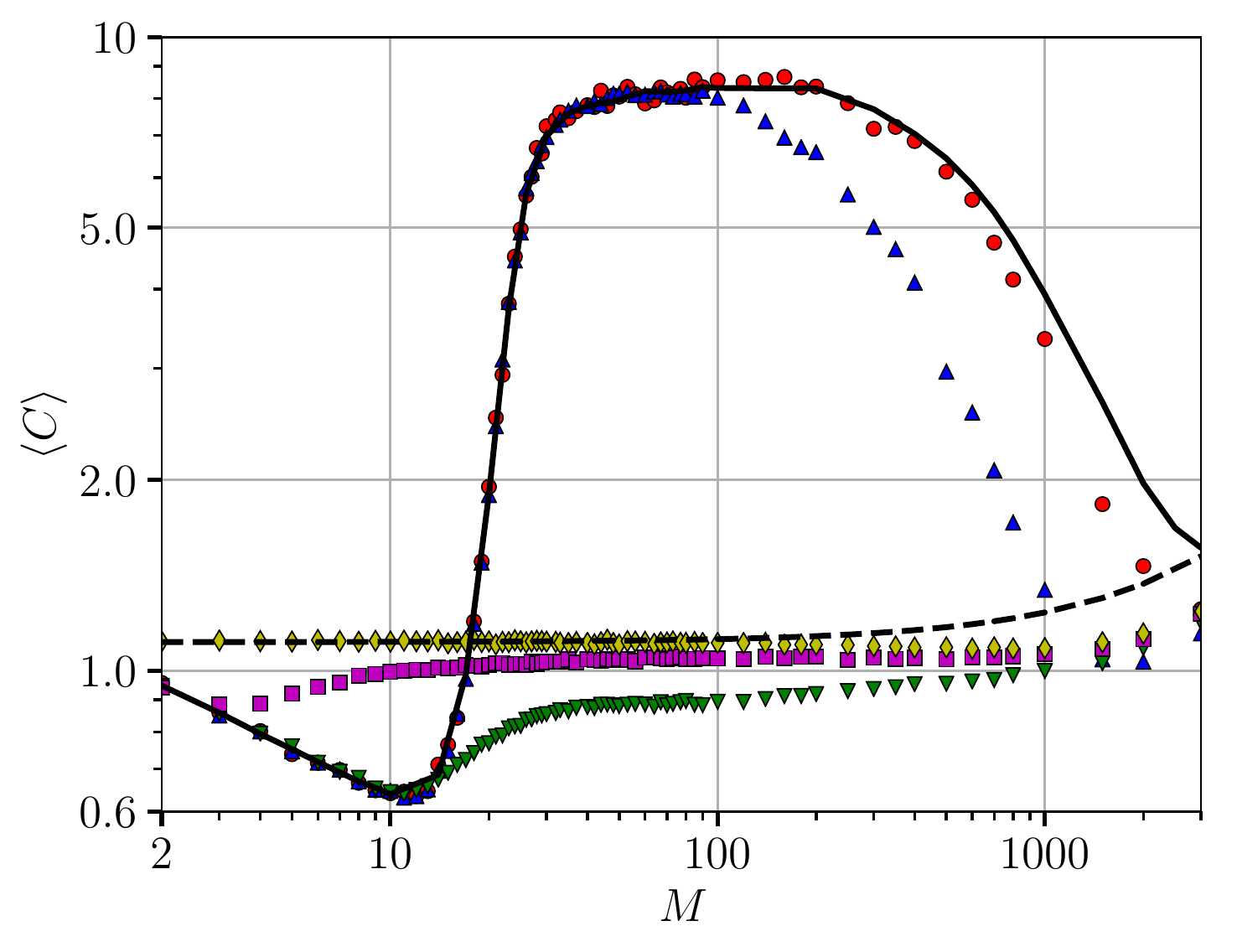}
\caption{Mean computational cost $\langle C \rangle $ as function of  the system size $M$ for mobile agents with step size $\delta = 5$. The  imitation probability is $p=0.5$ and the  radius of the influence neighborhood is $d = \alpha  d_0 $ with  $\alpha= 0.25 (\blacklozenge), 1 (\blacksquare), 2  (\blacktriangledown), 5 (\blacktriangle) $ and $10 (\CIRCLE)$.
 The solid curve is the result for the fully connected system and the dashed curve is the analytical  result for the independent search \cite{Fontanari_15}.
 The parameters of the rugged landscape  are $N=12$ and $K=4$.
 }  
\label{fig:5}  
\end{figure}

Figure \ref{fig:5}  reveals the intricate  interplay between the  parameter $\delta$, which specifies the length of the step   $\delta d_0$, and the parameter $\alpha$, which determines the radius of the influence neighborhood $\alpha d_0$. For large $\alpha$,   the computational cost is little affected by the mobility of the agents since  the odds that an agent becomes isolated and thus escapes the influence of  the local maxima is
negligible in this case. For small $\alpha$  the mobility is irrelevant too, as the agents remain isolated regardless of their wanderings on the square box. There is, however, a range of values of $\alpha$ where the mobility is very influential and produces antagonistic effects on the computational cost. For instance,  comparison with the lower panel of Fig.\ \ref{fig:2} indicates that the mobility increases the cost and hence is detrimental for $\alpha =1$, whereas  it decreases the cost and hence is beneficial  for $\alpha=2$. 

\begin{figure}
\centering  
 \includegraphics[width=0.48\textwidth]{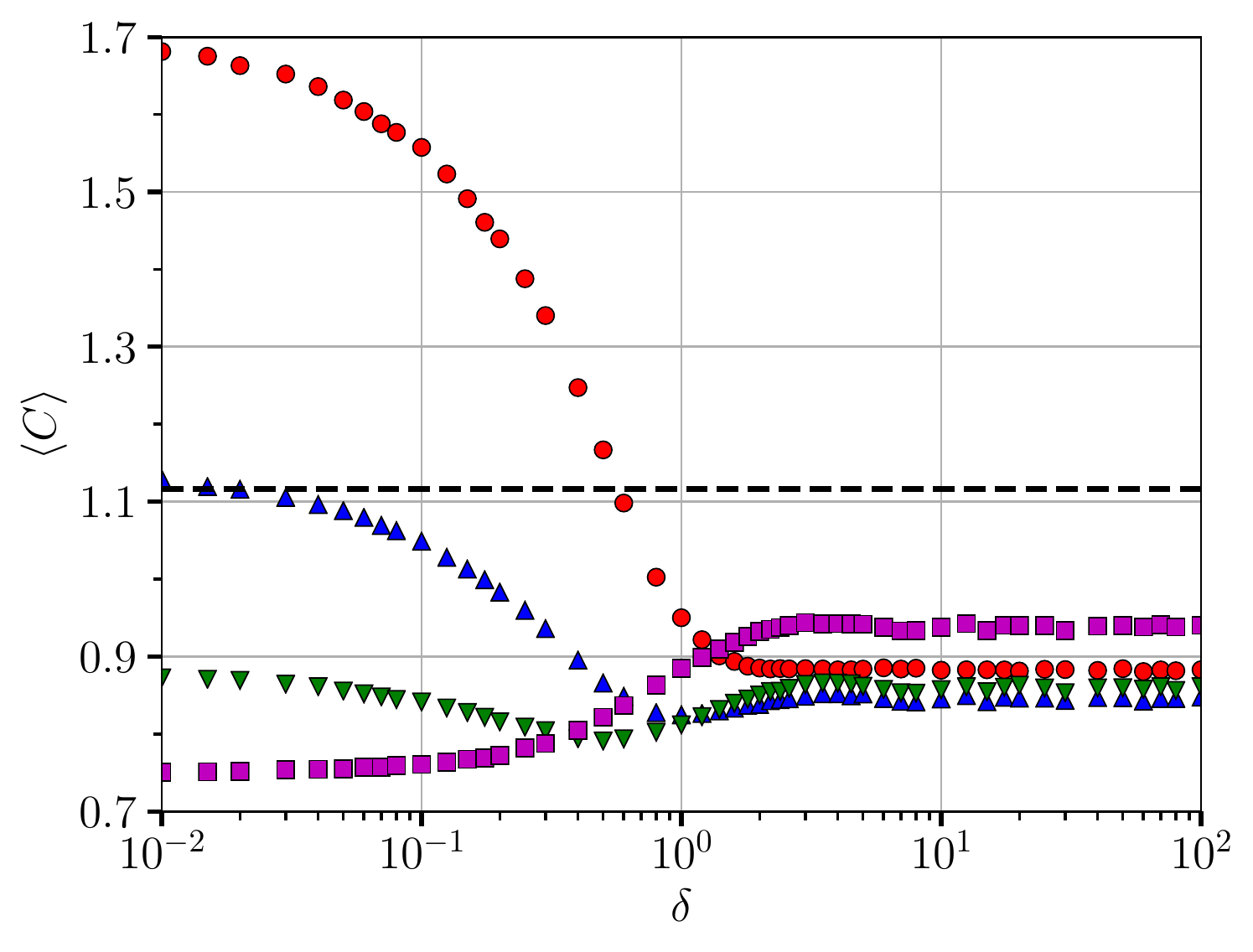}
\caption{Mean computational cost $\langle C \rangle $ as function of  the  step size $\delta$ for a  system of  size $M=53$. The  imitation probability is $p=0.5$ and the  radius of the influence neighborhood is $d = \alpha  d_0 $ with  $\alpha= 1.5 (\blacksquare),  1.8  (\blacktriangledown), 1.9 (\blacktriangle) $ and $2 (\CIRCLE)$. The horizontal dashed  line is the result for the independent search $\langle C \rangle \approx 1.12$.
 The parameters of the rugged landscape  are $N=12$ and $K=4$.
 }  
\label{fig:6}  
\end{figure}

The effect of mobility is more noticeable in Fig.\ \ref{fig:6} where we fix the system size
to $M=53$, which corresponds to the maximum of the computational cost in the lower panel of Fig.\ \ref{fig:4}, and vary the step size $\delta$ over several orders of magnitude. Since for this system size the trapping effects of the local maxima are maximized, moving the model agents  far away from their  clones is an efficient  way to mitigate the influence of those maxima, as seen in the case $\alpha=2$.  When the influence of the local maxima is already reduced due to the small influence neighborhoods of the agents, as  in the case of $\alpha=1.5$,  the mobility can actually help their dissemination over  the square box, resulting in the increase of the computational cost. In any event,  a  large step size $\delta$ guarantees that  the imitative search always  outperforms the independent search.

\begin{figure}
\centering  
 \subfigure{\includegraphics[width=0.48\textwidth]{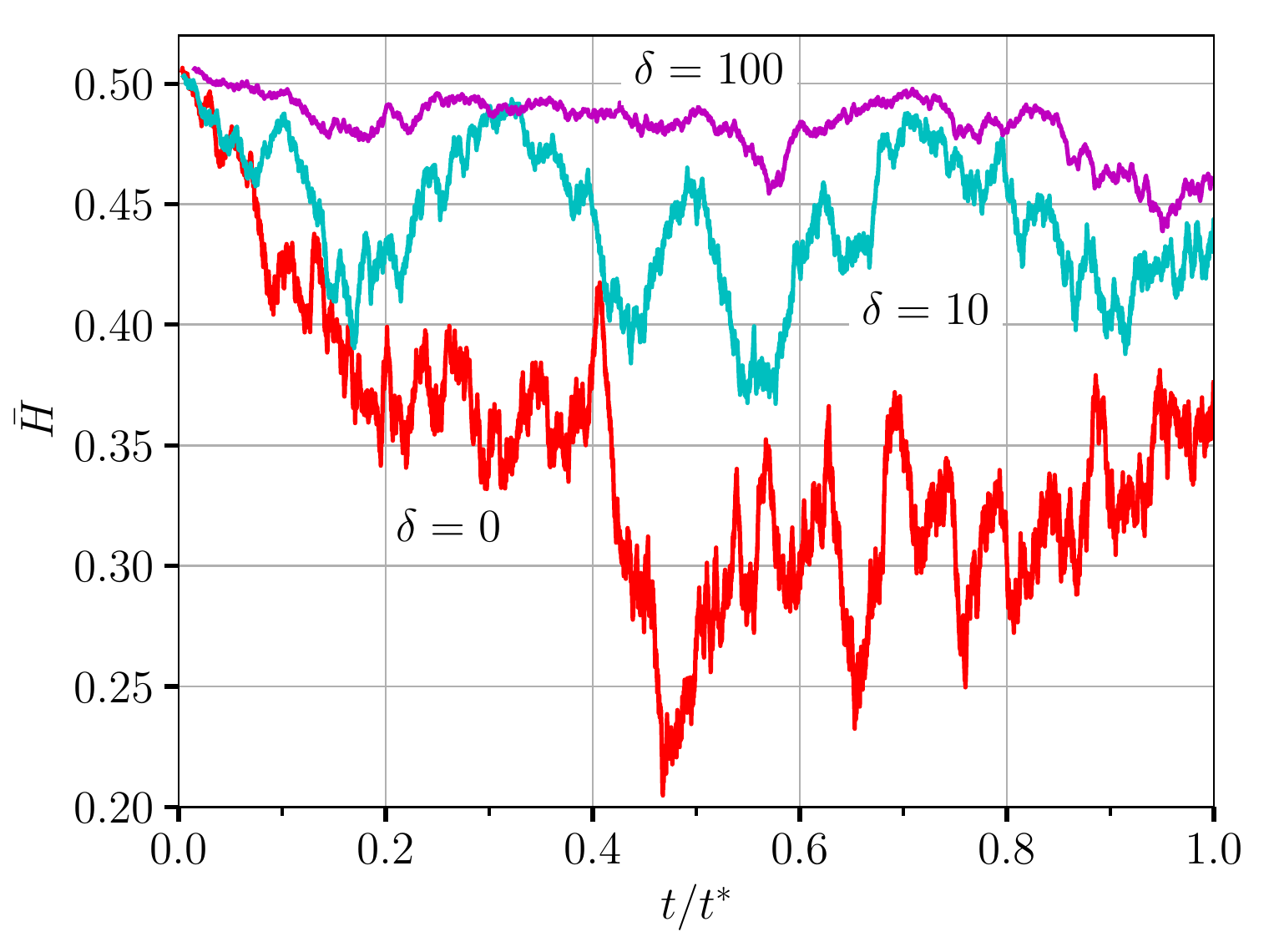}} 
 \subfigure{\includegraphics[width=0.48\textwidth]{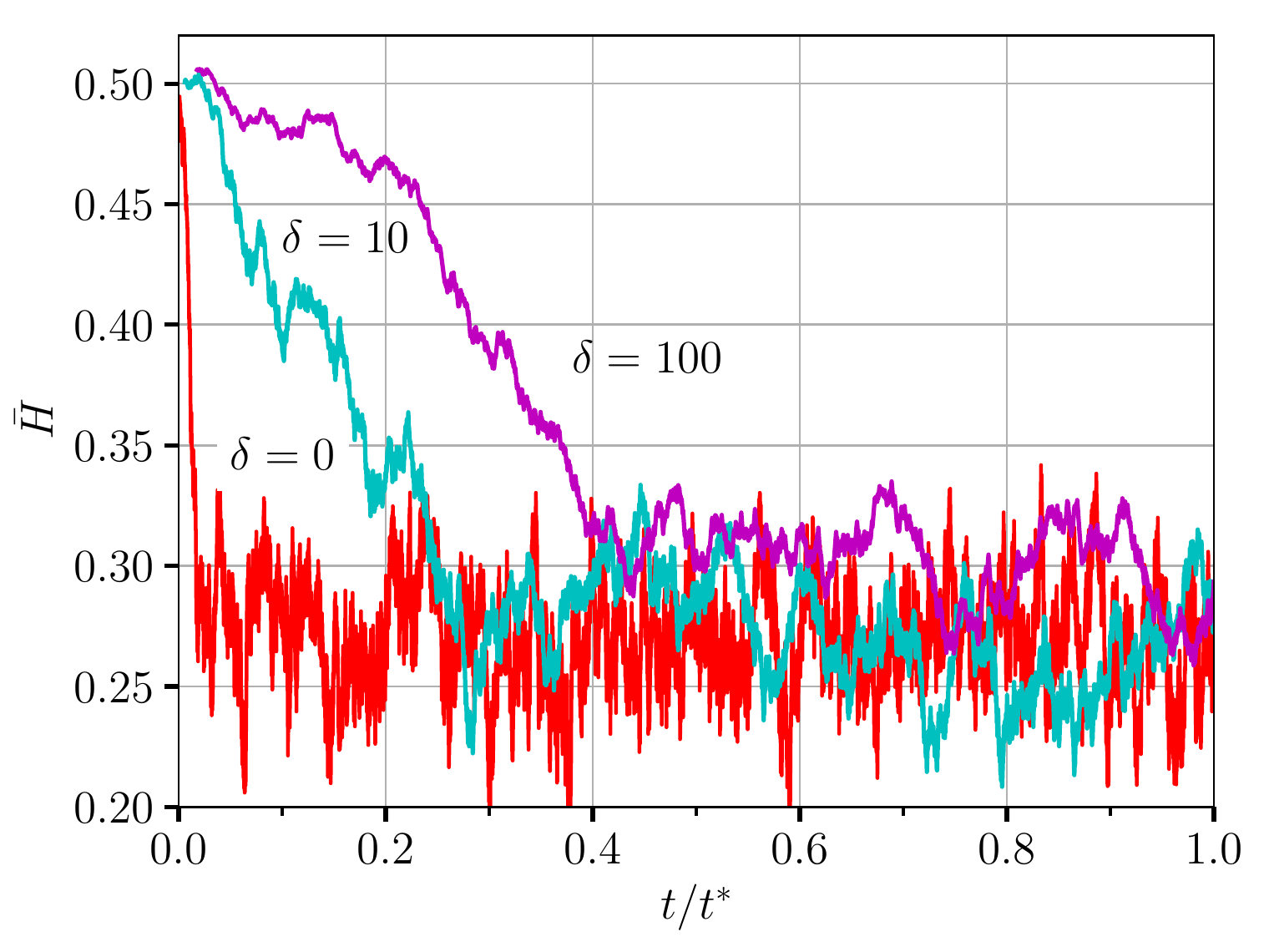}}
\caption{Time evolution of the mean pairwise Hamming distance $\bar{H}$  for   single runs of a system of  size $M=53$ and 
 step sizes $\delta=0$, $\delta=10$ and $\delta=100$ as indicated. The time $t$ is scaled by the halting time $t^*$ of each search.
 The  imitation probability is $p=0.5$ and the  radius of the influence neighborhood is $d = \alpha  d_0 $ with  $\alpha= 2$ (upper panel)
 and $\alpha= 4$ (lower panel).
 The parameters of the rugged landscape  are $N=12$ and $K=4$.
 }  
\label{fig:7}  
\end{figure}

To verify the soundness of our claim that the   high computational cost  is  caused by the  loss of diversity of the system when the search is trapped in  the local maxima,  we consider the time-dependence of  the mean pairwise distance between the $M$ strings in the system
\begin{equation}
\bar{H}= \frac{2}{M\left ( M-1 \right)} \sum_{k=1}^{M-1} \sum_ {l= k + 1}^M h \left ( \mathbf{x}^k, \mathbf{x}^l  \right ),
\end{equation}
 where
\begin{equation}
h \left ( \mathbf{x}^k, \mathbf{x}^l \right ) = \frac{1}{2} -  \frac{1}{2N} \sum_{i=1}^N  ( 1 - 2x_i^k  )  ( 1 - 2x_i^l  ) 
\end{equation}
is  the normalized Hamming distance between the bit strings $\mathbf{x}^k$ and $\mathbf{x}^l$ \cite{Hamming_50}.
The quantity $\bar{H}$  can be interpreted as follows:  if  we pick two strings at random, they will differ  by $N \bar{H}$ bits on average, and so  $\bar{H}$ measures the diversity of the strings in the system.  Figure \ref{fig:7}  shows the effect of the step size $\delta$ on the time evolution of  $\bar{H}$ for single runs of the imitative search. The panels show the results for  two  values of the radius of influence of the agents $\alpha d_0$, viz., $\alpha=2$ and $\alpha=4$. Since the initial strings are chosen randomly,  one has
$\bar{H} =0.5$ at $t=0$, which corresponds to the maximum  diversity. For $\alpha=2$, highly mobile agents  can maintain the high diversity of  the  system  during  the entire search, thus indicating  that the local maxima have little influence  on the computational cost in accord with   Fig.\ \ref{fig:4}. In the case of motionless agents ($\delta=0$), however,  the diversity decreases somewhat abruptly  in  the first half of the search, resulting in  a situation  where the strings differ from  each other by $2.5$ to $3.5$ bits on average, which  hints that the search is  stuck  in  local maxima. In the second half of the search, the diversity  increases slowly,  suggesting that a fraction of the agents managed to escape the local maximum traps and  found their way to the  global maximum.  For $\alpha=4$,   the trapping effects of the local maxima are greatly enhanced, as expected. Although  highly mobile agents can delay the fall into those traps, the search eventually gets stuck   resulting in the confinement  of a substantial fraction of the agents in the neighborhoods of the local maxima. 
These results thus support our claim that the loss
of diversity (i.e. Groupthink)  due to the trapping effects of the local maxima is the ultimate culprit for the high computational cost of the imitative search.

In our model, the exploration of the state space of the NK-fitness landscape halts at $t^*$ when one of the agents reaches the global maximum. If, however, the search is allowed to continue, then the agent that first found the global maximum will quickly attract the rest of the group to its neighborhood, similarly to what happens for the local maxima. The strength of the attraction decreases with increasing mobility as shown in the  lower panel of Fig. \ref{fig:7}. Hence a considerable number of agents  will be at  or in the close vicinity of  the global maximum after $t^*$, so a more stringent halting criterion will not significantly affect  the computational cost. 

A word is in order about an intriguing feature of the imitative search manifested in Figs.\ \ref{fig:2} and \ref{fig:4} for the rugged landscape ($K=4$), namely, the appearance of a shallower minimum of the computational cost for $M \approx 300$ and not too large step sizes. The explanation has to do with the  antagonistic effects of varying the system size. On the one hand,
increasing $M$ beyond the optimal size ($M \approx 12$)  allows the appearance of clones of the model string,  which  strengthens the attractivity of the local maxima and leads to  the sharp increase of the computational cost shown in those figures.  On the other hand, increasing $M$  makes the network sparser and delays the  flow of information over it, which reduces the influence of the local maxima.     To see this we calculate the average  path  length $\bar{l}$ of the influence networks, defined  as the average number of steps along the shortest paths for all
possible pairs of network nodes \cite{Newman_10}. This calculation is done  for systems with  $\alpha  \geq 2$ only, for which the shallower minimum appears (see Fig.\ \ref{fig:2}) and whose influence networks have a high average connectivity $\langle k \rangle  > 12$ (see Fig.\ \ref{fig:3}) so  we can guarantee that they  are  connected.
For the (connected) influence networks with fixed $d = \alpha d_0$ we find that $\bar{l}$ increases with $M^{0.47}$ as in the regular square lattice \cite{Vragovic_05} (it increases with $\ln M$ for Erd\H{o}s-R\'enyi graphs \cite{Albert_02}),  which justifies our claim that information flows slower for large $M$.  It is this  weakening of the influence of the local maxima that  leads to the shallower minimum of the computational cost exhibited in Figs.\ \ref{fig:2} and \ref{fig:4}. Lastly,
 further increase of $M$  leads to performance degradation due to the duplication of work as in the case of the smooth landscape. 

The antagonistic effects of increasing the system size on the dynamics of the search (creation of clones boosts the attractivity of the local maxima) and on the topology of the influence network (slow down of the flow of information in the system) are behind  the  nontrivial outcomes discussed in this section. In addition,  the position-fixed scenario ($\delta=0$) yields the best performance for the second-best  system size because for  sparse networks (see, e.g., the results for $\alpha=1.5$ in Fig.\ \ref{fig:6}) the mobility speeds up the  dissemination of the local maxima over  the square box.

Finally, we mention that we also considered different realizations of rugged landscapes with $K=4$, as well as landscapes with distinct degrees of epistasis,  and found that the same qualitative conclusions regarding the roles of the parameters $\alpha$ and $\delta$ hold true for every landscape realization.

 \section{Discussion}\label{sec:disc}
 
 It has long been known that  patterns of communication, i.e. who can communicate with whom, have a strong influence on the  problem-solving efficiency of groups \cite{Bavelas_50,Leavitt_51} (see \cite{Mason_12,Sebastian_17} for more recent contributions). These studies focused on  imposed or fixed  communication patterns, which is typical of  the military and industrial organizations,  thus excluding a priori  the possibility  of self-organization of the group members. A simple way to introduce flexibility on the patterns of communication is  to allow the agents to roam around an arena where they can interact with each other  if the distance between them is less   than a prespecified threshold. Within this roaming scenario,  we study  the performance of cooperative problem-solving systems that use imitative learning \cite{Fontanari_14} as the search  strategy  to find the global maxima of NK-fitness landscapes.
 
We find that for smooth landscapes, i.e. landscapes with a single maximum, mobility is always slightly detrimental to the  imitative search performance. Hence in the case the information exchanged among the agents  (i.e., their fitness values)   correlates  strongly with their distances to the global maximum,  the best strategy is to maintain and strengthen the local spatial correlations between the agents by keeping them fixed at their initial positions. However, for rugged landscapes, where the presence of local maxima uncouples the fitness values from the distances to the global maximum, imitation of   high fitness agents may lead to entrapment in the  local maxima. In this case,   mobility offers a  mechanism to circumvent those traps with the   guarantee of  always outperforming the independent search and reproducing the overall optimal performance achieved   by fully-connected small systems.

We stress that our study of the imitative  search does not seek to offer an alternative heuristic to tackle optimization problems. Rather, it seeks to assess quantitatively the potential of imitative learning as the underlying cooperative mechanism  of task-oriented groups.
In fact, since  finding the global maxima of NK landscapes  with $K>0$ is an NP-Complete problem \cite{Solow_00},  we do not expect that the imitative  search (or, for that matter, any other search strategy)  will reach those maxima much more rapidly
than the independent search. However,  finding the solution much more slowly than the independent  search, as observed for certain values of the group size $M$  (see, e.g.,  Fig. \ref{fig:5}), is a bad  omen for a search strategy. But this negative outcome is actually the main thrust of the imitative search since it is akin to a well-known maladaptive behavior associated to  social learning, namely, the Groupthink phenomenon 
that occurs when everyone in a group starts thinking alike \cite{Janis_82}.

Nevertheless, it is instructive to compare the imitative  search with the  evolutionary  algorithms  \cite{Back_96} as there are clear
similarities between those two heuristics. In particular,  flipping  a randomly chosen bit of the target string resembles the  mutation operator  of the evolutionary  algorithms, except that in those  algorithms mutation is an error of the reproduction process, whereas in the imitative search flipping a bit  at random and imitating the model string are mutually exclusive processes. In addition, imitation resembles  the 
crossover process of genetic algorithms, with the  caveat that the model agent is a mandatory parent in all mates but it contributes only  a single bit  to the offspring, which then replaces the other parent, namely, the target agent. More importantly, the bit passed to the offspring  is not random since it must be absent in the target string. This aspect highlights the fact that the imitative search models cultural, rather than genetic,  inheritance.

An appealing feature of the imitative search strategy as a model of human cooperation in  problem solving  is the emergence of Groupthink.  Real-life remedies for this
issue include the call for outside experts to share their viewpoints with the group members and the  leave of members to facilitate exposure to fresh ideas outside the influence of the group.  Both ventures involve the notion of mobility,  hence  our interest in finding  out whether mobility has similar beneficial effects on the imitative learning search too. However, the way we introduce mobility in the search, such that all members of the group can move  thus allowing  a group to break
 apart due to the motion of its members,  does not tally with the aforementioned remedies. A metapopulation approach where the group is composed of stable subgroups and the agents can migrate between them seems to offer a more suitable scenario to describe those situations.  Nevertheless,  our scenario offers
a first step in the  study of the effects of the flexibility on the patterns of communication and our finding that the random motion of agents  can avoid Groupthink  supports the common sense view that some kind of mobility is beneficial to group problem solving.

\bigskip

\acknowledgments
The research of JFF was  supported in part 
 by Grant No.\  2017/23288-0, Fun\-da\-\c{c}\~ao de Amparo \`a Pesquisa do Estado de S\~ao Paulo 
(FAPESP) and  by Grant No.\ 305058/2017-7, Conselho Nacional de Desenvolvimento 
Cient\'{\i}\-fi\-co e Tecnol\'ogico (CNPq).
SMR  was supported by grant  	15/17277-0, Fun\-da\-\c{c}\~ao de Amparo \`a Pesquisa do Estado de S\~ao Paulo 
(FAPESP). FAR acknowledges the Leverhulme Trust, CNPq (Grant No. 305940/2010-4) and FAPESP (Grants No. 2016/25682-5 and grants 2013/07375-0) for the financial support given to this research.

\appendix*
\section{}
 In this appendix we sketch a probabilistic description of the states of the agents evolving under the rules of the imitative search described in Section \ref{sec:model}.
 We begin by introducing some notation. The state of agent $k$ is represented by the binary string $\mathbf{x}^k = \left ( x_1^k, \ldots,x_N^k \right )$ with
 $x_i^k=0,1$;  $i=1,\ldots,N$ and $k=1,\ldots,M$. The set of agents, including agent $k$, in the influence neighborhood of agent $k$ is denoted by $\Omega_k$. 
 At time $t$, the model agent $m$ in  $\Omega_k$  is represented by the string $\mathbf{x}^{m}$ so that
 $ \mathcal{F} \left ( \mathbf{x}^{m} \right ) \geq \mathcal{F} \left ( \mathbf{x}^l  \right ) $ for  all $l \in \Omega_k$.  In addition, we let  $\mathbf{\tilde{x}}^{k_i} = \left ( x_1^k, \ldots, x_{i-1}^k, 1-x_i^k,  x_{i+1}^k, \ldots, x_N^k \right )$  represent  the 
 state of agent $k$ that differs from $\mathbf{x}^k$  solely at bit $i$. 
  
 Since in the elemental time interval $\Delta t=1/M$ application of the update rules   results always in the flipping of   one bit of the target agent, the state of agent $k$ will only change  when  it is chosen as the target agent, which happens with probability $1/M$. The probability $P_{t + \Delta t} \left ( \mathbf{x}^k  \right )$  that agent $k$ is represented by the string $\mathbf{x}^k$ at time $t + \Delta t$ is then written as 
\begin{eqnarray}\label{A1}
 P_{t + \Delta t} \left ( \mathbf{x}^k  \right )   &  = & \frac{1}{M} \left ( 1-p \right ) \frac{1}{N}  \sum_i^N  P_t \left ( \mathbf{\tilde{x}}^{k_i} \right ) \nonumber \\
&  & \mbox{} +  \frac{1}{M} p \frac{1}{N}  \sum_i^N  P_t \left ( \mathbf{\tilde{x}}^{k_i} \right ) \delta \left ( \mathbf{\tilde{x}}^{k_i},\mathbf{x}^{m} \right )
\nonumber \\
&  & \mbox{} +   \frac{1}{M} p \frac{1}{N} \sum_i^N  P_t \left ( \mathbf{\tilde{x}}^{k_i} \right ) \Lambda \left (  \mathbf{x}^k, \mathbf{\tilde{x}}^{k_i} ,\mathbf{x}^{m} \right ) \nonumber \\
&  & \mbox{} + \left ( 1 - \frac{1}{M} \right )  P_{t} \left ( \mathbf{x}^k  \right )
\end{eqnarray}
where 
\begin{equation}\label{A2}
\Lambda \left (  \mathbf{x}^k, \mathbf{\tilde{x}}^{k_i},\mathbf{x}^{m} \right  )   = \frac{\left [ 1 - \delta \left ( \mathbf{\tilde{x}}^{k_i},\mathbf{x}^{m} \right )\right ]
\delta \left ( x_i^k,x^m_i \right )}{\sum_j^N \left [  1 -  \delta \left ( \tilde{x}_j^{k_i},x^m_j \right ) \right ]},
\end{equation}
with
$\delta \left ( \mathbf{x},\mathbf{y} \right ) = 1$ if the binary strings  $\mathbf{x}$ and $\mathbf{y}$ are identical and 
$\delta \left ( \mathbf{x},\mathbf{y} \right ) = 0$, otherwise. We have also used the scalar  version of this function, namely, the Kronecker delta,
$\delta \left ( x , y \right ) = 1$ if $x=y$ and $0$, otherwise. 

The first term on the LHS of eq.\ (\ref{A1}) describes the random flipping of a bit, which   occurs with probability $1-p$, and accounts for the
possibility  that at time $t$  the state of  agent $k$ is $\mathbf{\tilde{x}}^{k_i}$ and that bit $i$ is flipped.  The second  term on the LHS of eq.\ (\ref{A1})
describes the imitation procedure, which  happens with probability $p$, for the special situation  where   agent $k$, whose state  is $\mathbf{\tilde{x}}^{k_i}$, is the model agent at time $t$.
We recall that in this case, the target agent flips a bit at random.  The third  term on the LHS  of eq.\ (\ref{A1}) describes the general imitation process when the target string differs from the model string.  As before, the state of  agent $k$ at time $t$ is  $\mathbf{\tilde{x}}^{k_i}$, but now bit $i$  must be copied from the model string.  The probability  of this event is given by the reciprocal of the number of different bits in strings  $\mathbf{\tilde{x}}^{k_i}$ and $\mathbf{x}^{m}$, i.e., the reciprocal of the Hamming distance between these two strings. We note that $\mathbf{\tilde{x}}^{k_i}$ and $\mathbf{x}^{m}$  differ  by one bit at least, namely, bit $i$, so the denominator in eq.\ (\ref{A2}) never vanishes. Finally, the fourth term on the LHS of eq.\ (\ref{A1}) accounts for the case that the state of agent $k$ at time $t$ is $\mathbf{x}^k $ and that this agent is not chosen as the target agent.

Using $\Delta t = 1/M$ we can rewrite eq.\ (\ref{A1}) as
\begin{eqnarray}\label{A3}
 \frac{\Delta P_{t} \left ( \mathbf{x}^k  \right )}{\Delta t}   &  = &   - P_{t} \left ( \mathbf{x}^k  \right ) +  \frac{1-p}{N}  \sum_i^N  P_t \left ( \mathbf{\tilde{x}}^{k_i} \right ) \nonumber \\
&  & \mbox{} +   \frac{p}{N}  \sum_i^N  P_t \left ( \mathbf{\tilde{x}}^{k_i} \right ) \delta \left (\mathbf{\tilde{x}}^{k_i},\mathbf{x}^{m} \right )
\nonumber \\
&  & \mbox{} +    \frac{p}{N} \sum_i^N  P_t \left (\mathbf{\tilde{x}}^{k_i} \right ) \Lambda \left (  \mathbf{x}^k, \mathbf{\tilde{x}}^{k_i},\mathbf{x}^{m} \right )
\end{eqnarray}
where $\Delta P_{t} \left ( \mathbf{x}^k  \right ) =  P_{t + \Delta t} \left ( \mathbf{x}^k  \right ) - P_{t} \left ( \mathbf{x}^k  \right )$. Hence the continuous-time limit is obtained  for  $M \to \infty$.  To take into account the fact that the search stops at the global maximum $\mathbf{x}^{g}$, we  need only to add the proviso  that $\mathbf{x}^{g}$ does not  appear in the argument of $P_t$ on the  LHS of eq.\ (\ref{A3}),  i.e., $\mathbf{x}^{g}$ is a perfect trap.  In addition to the  high dimensionality of the
state variables, the difficulty to iterate  eq.\ (\ref{A3})  is  due to the determination of the model string $\mathbf{x}^{m}$, which is actually the term that couples the $M$ agents in the system. In particular, $\mathbf{x}^{m}$ is the string   that maximizes $\mathcal{F} \left ( \mathbf{x}^l \right )$  for $l \in \Omega_k$ with the constraints that $ P_t \left ( \mathbf{x}^m \right ) > 0$ and  $\mathbf{x}^m \neq \mathbf{x}^{g}$, since  the global maximum is never a model string in the imitative search.   Finally, we note that in the case of mobile agents we need to include a time dependence on the  influence neighborhoods $\Omega_k$ and specify how they change with time.

\end{document}